\def\oii{{[O\thinspace{\sc ii}]}}
\def\p.{^{\prime\prime}\kern-2.1mm .\kern+.6mm}

\def\kms{\ifmmode{\,\hbox{km}\,s^{-1}}\else {\rm\,km\,s$^{-1}$}\fi}

\def\kmsm{{\rm\,km\,s^{-1}\,Mpc^{-1}}}

\def\hmpc{\ifmmode{h^{-1}\,\hbox{Mpc}}\else{$h^{-1}$\thinspace Mpc}\fi}
\def\hkpc{\ifmmode{h^{-1}\,\hbox{kpc}}\else{$h^{-1}$\thinspace kpc}\fi}
\def\eg{{\it e.g.}~}

\def\et{{\it et~al.}~}

\def\sigp{\ifmmode{\sigma_p}\else {$\sigma_p$}\fi}
\def\sig1{\ifmmode{\sigma_1}\else {$\sigma_1$}\fi}
\def\r200{\ifmmode{r_{200}}\else {$r_{200}$}\fi}

\documentstyle[11pt,aaspp4,flushrt]{article}
\tighten

\slugcomment{DRAFT: \today}

\begin{document}

\title
{Faint $K$-Selected Galaxy Correlations and Clustering Evolution}

\author
{
R.~G.~Carlberg,\altaffilmark{1}
Lennox L. Cowie,\altaffilmark{2,3,4} 
Antoinette Songaila,\altaffilmark{2,3}
and Esther M. Hu\altaffilmark{2,3,4}}

\altaffiltext{1}{Department of Astronomy, University of Toronto, 
	Toronto ON, M5S~3H8 Canada\\ email: carlberg@astro.utoronto.ca}
\altaffiltext{2}{Institute for Astronomy, University of Hawaii, 
	2680 Woodlawn Dr.,
	  Honolulu, HI 96822\\  email: acowie, cowie \& hu@ifa.hawaii.edu}
\altaffiltext{3}{Visiting Astronomer, W. M. Keck Observatory, jointly
  operated by the California Institute of Technology and the University of
  California.}
\altaffiltext{4}{Visiting Astronomer, Canada-France-Hawaii Telescope,
  operated by the National Research Council of Canada, the Centre National
  de la Recherche Scientifique of France, and the University of Hawaii.}


\begin{abstract}
Angular and spatial correlations are measured for $K$-band--selected
galaxies, 248 having redshifts, 54 with $z>1$, in two patches of
combined area $\simeq27$~arcmin$^2$.  The angular correlation for
$K\le21.5$ mag is $\omega(\theta)\simeq
(\theta/1.4\pm0.19^{\prime\prime}e^{\pm0.1})^{-0.8}$.  From the
redshift sample we find that the real-space correlation, calculated
with $q_0=0.1$, of $M_K\le-23.5$ mag galaxies (k-corrected) is $\xi(r)
= (r/2.9e^{\pm0.12}\hmpc)^{-1.8}$ at a mean $z\simeq 0.34$,
$(r/2.0e^{\pm0.15}\hmpc)^{-1.8}$ at $z\simeq 0.62$,
$(r/1.4e^{\pm0.15}\hmpc)^{-1.8}$ at $z\simeq 0.97$, and
$(r/1.0e^{\pm0.2}\hmpc)^{-1.8}$ at $z\simeq 1.39$, the last being a
formal upper limit for a blue-biased sample.  In general, these are
more correlated than optically selected samples in the same redshift
ranges.  Over the interval $0.3\le z\le0.9$ galaxies with red
rest-frame colors, $(U-K)_0>2$ $AB$ mag, have
$\xi(r)\simeq(r/2.4e^{\pm0.14}\hmpc)^{-1.8}$ whereas bluer galaxies,
which have a mean $B$ of 23.7 mag and mean \oii\ equivalent width
$W_{eq} = 41$~\AA, are very weakly correlated, with
$\xi(r)\simeq(r/0.9e^{\pm0.22}\hmpc)^{-1.8}$.  For our measured growth
rate of clustering, this blue population, if non-merging, can grow
only into a low-redshift population less luminous than $0.4L_\ast$.
The cross-correlation of low- and high-luminosity galaxies at
$z\simeq0.6$ appears to have an excess in the correlation amplitude
within 100
\hkpc.  The slow redshift evolution is consistent with these galaxies tracing
the mass clustering in low density, $\Omega\simeq 0.2$, relatively
unbiased, $\sigma_8\simeq0.8$, universe, but cannot yet exclude other
possibilities.\end{abstract}

\clearpage
\section{Introduction}

N-body simulations give reliable predictions for the redshift
dependence of the two-point correlation function of the density field,
$\xi(r|z)$, as a function of $\Omega$.  A convenient power-law
parameterization to describe the evolving correlation function of
galaxies is (\cite{gp,ks})
\begin{equation}
\xi(r|z)=\left({r\over r_0}\right)^{-\gamma} (1+z)^{-(3+\epsilon)},
\label{eq:def}
\end{equation}
where the lengths $r$ are measured in physical (proper) co-ordinates.
For this double power-law approximation the predicted evolution of clustering 
in the mass field is faster for $\Omega=1$
($\epsilon=1.0\pm0.1$) than it is for low-$\Omega$ values, for instance
$\epsilon=0.2\pm0.1$ for $\Omega=0.2$ (\cite{ccc}).  Therefore,
measurement of the redshift evolution of clustering can be used to test the
gravitational instability theory of structure formation and the relation
of galaxy clustering to dark matter clustering, and provides a
constraint on $\Omega$. Knowledge of these quantities enables 
predictions of the distribution of assembly times of dark halos, which
on the relatively small scales investigated here is of great interest
for the mass evolution of galaxies.

At present, observational measures of clustering evolution are
uncertain simply due to the difficulties of assembling large samples
of faint galaxies with sufficient sky coverage to give a
statistically representative sample. At low redshift the form of
nonlinear galaxy clustering is accurately established
(\eg\ \cite{dp,apm,lcrs_lin,lcrs_tucker}), with a basic
characterization of its dependencies on galaxy color, luminosity, and
morphology.  At higher redshifts the clustering is only now being
directly measured (\cite{cfrs,chuck}), although the small fields leave
concerns that field-to-field variations are not yet well controlled.

The galaxy luminosity function and its color dependence evolve
substantially over the redshift 0 to 1 interval
(\cite{cfrs_lf,autofib,cshc,cnoc_lf}).  Differential luminosity
evolution of blue and red galaxies, in which the blue galaxies are
less correlated at low redshift, can cause the apparent correlation of
a magnitude-limited sample to change faster than either of the two
underlying populations are changing. In this paper we report the
clustering properties of a very deep redshift survey selected in the 
$K$ band.  A near-IR selected survey has the enormous advantage that both
k-corrections and the evolutionary corrections are small, allowing
galaxy luminosities to be identified with total stellar mass with
reasonable confidence.  The Hawaii $K$-band survey (\cite{cshc})
with a couple hundred galaxy redshifts, is large enough to be
useful for correlation studies.  Furthermore this survey contains
galaxies up to a redshift of 2.19, which provides a fairly large
redshift baseline over which correlation changes can be measured.

The next section summarizes the sample properties.  Measures of the
angular correlation are given in Section 3 and of the real space
correlation function in Section 4. In Section 5 the 
correlations of red and blue galaxies and of low- and high-luminosity
galaxies are compared.  Section 6 discusses the redshift evolution of
galaxy correlations and compares the available data to various model
predictions. Section 7 summarizes our results. All measurements in
this paper assume $H_0=100\ h^{-1} \kmsm$ and $q_0=0.1$. It should be noted
that correlation amplitudes at $z\simeq 1$ are reduced by
about 30\% for $q_0=0.5$.

\section{The Hawaii $K$-Band Sample}

The Hawaii $K$-selected redshift survey of two fields constitutes a
nearly complete sample down to $K=20$, $I=23$, and $B=24.5$ mag.  The
survey is described in detail elsewhere (\cite{cshc}) although this
analysis uses 20 new redshifts that have been recently obtained to
complete the $B$ selected subsample. The sample of 248 galaxies with
redshifts in an area of about 27 square arcminutes compares favourably
with the moderate redshift CNOC sample, about 200 galaxies in a single
field covering 221 square arcminutes (\cite{chuck}), and with the 591
galaxies covering about 71 square arcminutes (a result of the high
sampling rate in $0.5\arcmin\times9.4\arcmin$ strips) in the CFRS
study (\cite{cfrs}).  On the other hand, the redshifts here extend all
the way from 0.08 to 2.19, with 80\% between 0.28 and 1.39. The number
in any one redshift interval is sufficient to make useful correlation
measurements (Figure~\ref{fig:nz}).  All magnitudes and colors used in
this paper are k-corrected. The sky positions of the objects are
plotted in Figure~\ref{fig:xy}, where open squares are galaxies,
crosses represent stars, and objects that were not observed or which
lack secure redshift identifications are shown with triangles.

The galaxy redshifts, plotted against the projected physical distance
in the RA direction from the field center, are shown in
Figure~\ref{fig:pie}. The symbol area is proportional to the
luminosity of the galaxy.  The sample is known to be somewhat
incomplete for the redder galaxies at the faintest magnitudes, which
are generally expected to be galaxies beyond redshift one with low
star formation rates.  Incomplete samples, provided that they have no
spatial bias, do not pose a problem for correlation studies, provided
that the unclustered background distribution is generated from a
smoothed version of the observed redshift distribution. We will
approximate the observed redshift distributions as being constant over
the various subranges of interest, which tests indicate to be an
adequate approximation for these data.

\section{The Angular Correlation Function}

The angular correlation function is estimated as $\omega(\theta) =
(DD-2DR+RR)/RR$ (\cite{ls}) where in a given range of angles $DD$ is
the number of data pairs, $DR$ is the number of pairs between the data
sample and a uniform random sample. This estimator is particularly
useful when the clustering amplitude is significantly less than unity,
as it is here.  The resulting angular correlation function for the
full photometric sample, $K\le 21.5$ mag, is shown in
Figure~\ref{fig:aw}. The errors are assigned using the bootstrap
method (\cite{et}) with 100 resamplings.  The angular galaxy
correlations are diluted by the uncorrelated foreground stars, because
the full photometric sample is used, which requires that the
correlation amplitude be corrected upward by a factor of
$(1-f_\ast)^{-2}$, where $f_\ast$ is the fraction of objects which are
stars in the two combined fields. We estimate that $f_\ast=0.25$,
based on the stellar fraction of the spectroscopically identified
sample that are stars.  The uncorrected correlation function is
$\omega(\theta)\simeq (\theta/0\p.75)^{-0.8}$, from which we find
that the correlation angle of the galaxies is $\theta_0=1\p.4\pm0\p.2$
arcseconds, where the error is the internal error of the fit.  Several
recent studies in the optical region have found correlation angles of
$\theta_0\simeq1\arcsec$ at $R=23.5$ mag (\cite{hl}), $\theta_0\simeq
0\p.3$ at $I=22.5$ mag (\cite{lp}) and in a much fainter sample to
$r=26$ mag, $\theta_0\simeq 0\p.06$ (\cite{bsm}).  Since our mean
$\langle {I-K} \rangle\simeq 2.5$ we note that the angular
correlation that we observe is substantially larger than for a
comparable $I$- or $R$-band sample. This is a consequence of the
$K$-band sample's being fundamentally more correlated than optical
samples; we provide direct evidence for this below.

In a survey done with a multi-object spectrograph there is the
possibility that the angular correlation of the galaxies with
redshifts is biased relative to the parent sample as a result of
instrumental constraints for the selection of objects for
spectroscopy. The simplest test for a selection effect of this type is
to measure the ratio of $DD$ pairs as a function of separation in the
photometric and redshift sample as normalized to the total numbers in
the two samples. We find that there is a 10\% reduction in the number
of pairs within about $10\arcsec$, and a 5\% overselection at
separations around $30\arcsec$, diminishing with increasing
angle. This bias is generally less than the statistical errors, so we
do not apply any geometric corrections in the correlation
measurements.

\section{Real Space Correlations}

The velocity precision in this redshift survey is not adequate to
allow a measurement of the redshift space correlation function at
small scales.  However, we can measure the projected real space
correlation function,
\begin{equation}
w_p(r_p) = \int^\infty_{-\infty} \xi(\sqrt{r_p^2+y^2})\, dy, 
\label{eq:wp}
\end{equation}
(\cite{lss,dp}). The primary drawback of this estimator is that it
averages over long cylindrical shells in redshift space, so it has a
very broad window function (\cite{peacock}).

The real space correlation is derived from a power-law fit to the
projected correlation function, Eq.~\ref{eq:wp}, calculated from the
galaxy sample. Operationally, $w_p(r_p)$ is the integral over $r_v$ of
the 2D correlation function $\xi(r_p,r_v)$, where $r_p$ and $r_v$ are
the proper separation of galaxy pairs in the projected and redshift
directions, respectively.  Because our sample is not very large the
errors are dominated by the small number statistics, rather than any
complications of the estimator. We estimate $\xi(r_p,r_v)$ as
$DD/DR-1$, in the $(r_p,r_v)$ co-ordinates.  The random sample is 5 to
10$\times10^4$ points.  The sum over the $r_v$ co-ordinate is cut off
at a practical range of 10\hmpc\ in proper co-ordinates at the
redshift of the object (\cite{chuck}). This value was selected from a
range of trial values as being approximately the optimal value to
maximize the signal-to-noise.  This cutoff distance is sufficiently
large that the correction to the integral for the correlation beyond
the cutoff is everywhere less than about 20\%, and is ignored. The
redshift distribution of the random sample is generated assuming that
the unclustered redshift distribution is constant over the various
redshift and luminosity ranges. Altering the redshift ranges shows
this to be an entirely adequate approximation for these data. The
errors are bootstrap estimates, reduced by a factor of $\sqrt{3}$
(\cite{mjb}).

The $w_p(r_p)$ measured over the redshift ranges 0.2--0.4, 0.4--0.8,
0.8--1.2, and 1.2--1.6 are shown in Figure~\ref{fig:wp}. The galaxies
have minimum luminosities of $M_K\le -23.5$ mag in the three higher
redshift bins and $M_K\le -21.5$ in the lower redshift range, to
increase the sample size.  A rest-frame luminosity of $M_K= -23.5$ mag
is about half of $L_\ast$.

The $w_p(r_p)$ are fitted
to a power-law correlation function,
\begin{equation}
w_p(r_p) = r_0^\gamma {{\Gamma({1\over2})\Gamma({{\gamma-1}\over2})}
	\over \Gamma({\gamma\over2})} r_p^{1-\gamma},
\label{eq:wpc}
\end{equation}
(\cite{lss}).  The Gamma function factor is 3.68 for $\gamma=1.8$.
Over the redshift range $0.2\le z \le 0.4$ we find
$\xi(r)\simeq(r/2.9e^{\pm0.12}\hmpc)^{-1.8}$ at a mean $z\simeq 0.34$.
For the $0.4 \le z \le 0.8$ range we find
$\xi(r)\simeq(r/2.0e^{\pm0.15}\hmpc)^{-1.8}$ at a mean $z\simeq 0.62$,
and for $0.8\le z \le 1.2$,
$\xi(r)\simeq(r/1.4e^{\pm0.15}\hmpc)^{-1.8}$ at a mean $z\simeq0.97$.
Using wider projected radius bins we find that for the galaxies with
redshifts 1.2--1.6, $\xi(r)\simeq(r/1.0e^{\pm0.2}\hmpc)^{-1.8}$ at a
mean $z\simeq1.39$.  This high redshift correlation is formally an
upper limit on the basis of the bootstrap error estimates, although
the Poisson errors do indicate a significant measurement. We will
accept the result as an indicative measurement, because it remains
unclear how best to estimate the errors.  The galaxies in this highest
redshift bin are dominated by the faint $B$ subsample and are not
complete in the $K$ subsample This selection bias likely leads to an
underestimate of the correlation at this redshift if the enhanced
correlation of red-selected galaxies over blue-selected ones seen at
lower redshift is present at this redshift.

The decrease in the correlation length with increasing redshift is
significant within these data although the quoted errors are purely
the internal errors of the fit and do not account for field-to-field
differences beyond these two patches. The fitted $\epsilon
=0.2\pm0.5$, based on our $K$-band sample alone.

\section{Correlation Dependence on Color and Luminosity}

The color dependence of faint galaxy clustering is an important clue
to the formation mechanisms of galaxies. Higher mass galaxies are
expected to be more correlated at formation and subsequently
environmental effects can modify the correlations (\eg\
\cite{apm}).  
In Figure~\ref{fig:wc} the $w_p(r_p)$ are shown for the $0.3\le z \le
0.9$ galaxies with colors redder or bluer than $(U-K)_0=2$ mag
(k-corrected, rest-frame ratios between the flux $f_\nu$ at 3500\AA\
and at 21000\AA\ on the AB magnitude system, see \cite{cshc}). The red
galaxies, with $r_0=2.4e^{\pm0.14}\hmpc$, are about 5 times (with
about a 50\% error) more strongly clustered than the blue objects,
which have $r_0=0.9e^{\pm0.22}\hmpc$, both at fixed $\gamma=1.8$.  The
red galaxies appear to have a substantially steeper correlation slope,
$\gamma$, than our adopted value of 1.8, although a survey with larger
sky area is needed to assess the slope.  The average $M_K=-22.5$ mag
for the blue sample, whereas it is $M_K=-23.9$ mag for the red sample,
which is quite a small luminosity difference for the large correlation
difference (\cite{apm,lcrs_tucker}).  The blue population, at a mean
$B$ of 23.7 mag, has strong \oii\ lines, with a mean $W_{eq}$ of
41\AA. Therefore, these are very likely to be substantially brightened
relative to their intrinsic luminosities at a more normal star
formation rate.  The blue galaxy population must largely disappear
from the normal galaxy population at low redshift, since its
correlation length would grow to only about 2 \hmpc\, given our
estimated $\epsilon\simeq0.2$. Within the luminosity range explored in
the low redshift APM survey, no population is this weakly correlated
(\cite{apm}) although the low luminosity, $M_b=-18.6$, late type
galaxies have $r_0=2.9\pm0.4\hmpc$. We conclude that these high
redshift faint blue galaxies cannot become a significant component of
the low redshift galaxy population above luminosities of $0.4L_\ast$.
It is possible that these objects merge with higher luminosity
galaxies to drive their evolution.

The clustering of lower luminosity galaxies is weaker than that of
high luminosity galaxies.  For $0.2 \le z \le 0.4$ the galaxies with
$M_K \ge -21.5$ have $r_0=1.8e^{\pm0.2}\hmpc$ and for $0.4 \le z \le
0.8$ those with $M_K \ge -23.5$ have $r_0=1.1e^{\pm0.2}\hmpc$.  In
both these redshift ranges, the higher-luminosity galaxies are more
correlated that the lower luminosity ones.  The luminosity difference
is about 2.5 magnitudes.  It is quite unlikely that the high-redshift,
high-luminosity, weakly-correlated population could have evolved into
a weakly-clustered, low-luminosity, low-redshift population. First, it
requires that the clustering is being reduced in physical co-ordinates
with time. Second, about 4 magnitudes of fading per unit redshift is
required, whereas a much slower rate of luminosity density evolution
of the population average is observed (\cite{cfrs_j}). Third, it would
require that the high-luminosity galaxies at lower redshift originate
from some unobserved high redshift population --- an effect which
$K$-band observations minimize through their relatively small redshift
corrections. On the other hand, this discussion does emphasize that a
more precise definition of comparable galaxy samples at low and high
redshift would be very desirable, ideally done based on the intrinsic
properties of the galaxies themselves.

The cross-correlation of low-luminosity galaxies, $M_K\ge -23.5$ mag,
with high-luminosity galaxies, $M_K \le -23.5$ mag, is shown in
Figure~\ref{fig:w2} for the $0.3\le z \le 0.9$ range (expanded
slightly to boost the sample size). Although the sample is smaller
than is really desirable it shows quite intriguing how much more
strongly the low luminosity galaxies cluster to high luminosity
``hosts'' within 100\hkpc, beyond which the cross-correlation drops to
a value similar to the field cross-correlation of low-luminosity
galaxies. The effect is directly visible in the redshift diagrams of
Figure~\ref{fig:pie}.  The enhanced cross-correlation of close pairs
is not primarily the result of an enhanced star formation which would
raise the optical band luminosity between 1 and 2 magnitudes, and
hence would increase the numbers above some flux limit. The lower
luminosity galaxies have substantially stronger \oii; however the
expected accompanying increase in the $K$ luminosity is
small. Consequently these data favour a genuine increase in close
pairs at small separations over the power-law correlation.

\section{Evolution of Galaxy Correlations}

To follow the clustering amplitude as a function of redshift we need
to compare the results of a variety of redshift surveys, having
varying $r_0$ and $\gamma$.  The quantity $r_0^\gamma$, is a useful
measure of the clustering amplitude, which can be interpreted either
as the amplitude at 1 \hmpc, or, given that these quantities are
normally the results of a fit to data over a range of scales,
$r_0^\gamma$ is a measure of the average correlation within a fixed
proper volume.

At low redshift there are several recent measurements of clustering
with a range of sample definitions.  The $b_J$-selected APM survey
finds $r_0=5.1\pm0.2\hmpc$ and $\gamma=1.71\pm0.05$ giving $r_0^\gamma
= 16.2\pm2.5$ at $z\simeq0.06$ (\cite{apm}). The $R$-selected LCRS
finds $r_0=5.0\pm0.14\hmpc$ and $\gamma=1.79\pm0.04$ giving
$r_0^\gamma=17.8\pm2.2$ at a mean $z\simeq0.1$ (\cite{lcrs_lin}). The
IRAS-selected correlation function is $r_0=3.76\pm0.2\hmpc$ and
$\gamma=1.66\pm0.11$ (\cite{iras}) for $r_0^\gamma=9.0\pm2.6$. There
is no large $K$-band selected redshift survey at low redshift; however
the angular correlation of bright K-selected galaxies is measured,
from which we estimate $r_0^\gamma= 27.5$ and 13.8 at $z=0.13$ and
$0.23$, respectively, with $\sim$20\% systematic errors (\cite{bgfs}
referred to as BFFS).  A preliminary measurement in the LCRS survey finds that
the red galaxies have a correlation comparable to these $K$-band
results (\cite{tucker}).

At higher redshift we have the results of this paper, the $r$-selected
CNOC field survey at a mean redshift of 0.36 (\cite{chuck} and work in
progress), which gives $r_0^\gamma\simeq9.6$ for the red half of the
sample and 7.5 for the blue half. The CFRS results (\cite{cfrs})
provide correlation estimates over the redshift range 0.2-0.9.  They
find $\xi(r)=(r/2.4\pm0.17\hmpc)^{-1.64}$ for $0.2\le z \le 0.5$,
$\xi(r)=(r/1.4\pm0.19\hmpc)^{-1.64}$ for $0.5\le z \le 0.75$, and
$\xi(r)=(r/1.4\pm0.20\hmpc)^{-1.64}$ for $0.75\le z \le 1$, where we
have adjusted their correlation lengths for $q_0=0.5$ to our $q_0=0.1$
using their formula.  The CFRS data indicate $\gamma=1.64$, which
would be a relatively poor description of the $K$-selected data here.
As a direct comparison with the CFRS measurements, we define a
complete subsample limited at $I=22.5$ mag from our data.  We find
that $r_0=1.9e^{\pm0.19}\hmpc$ for $0.3
\le z \le 0.9$, which is statistically
identical to the CFRS measurement of $1.8\pm0.18\hmpc$ (adjusted to
$q_0=0.1$) over the $0\le z \le 1$ range.

The correlation amplitudes are plotted against redshift in
Figure~\ref{fig:ev}. For comparison the similarly measured
correlations from n-body simulations (\cite{ccc}) are also shown, all
scaled with a linear multiplicative factor roughly to fit the LCRS
correlation measurement. The fitted $\gamma$ values from these
simulations are in the range 1.8-2.0, which is compatible with the
$K$-band sample but not the shallower slopes usually seen in optically
selected samples.  The scaling factors are in the range of 0.55 to
0.70, which is a little larger than desirable (ideally one would do
specially matched simulations), however, these factors are small
compared to the factor of 20 or so in the evolution of the correlation
functions.

There are two conclusions to be drawn from Figure~\ref{fig:ev}.
First, optically selected galaxies appear to always be significantly
less correlated than $K$-selected galaxies, the difference being
typically about a factor of two in the amplitude.  Second, for the
$K$-band and red-selected samples, which should be least sensitive to
galaxy population evolution, we see that the evolution of galaxy
clustering is reasonably well described by an $\Omega=0.2$
model. However, the amplitude measured in the n-body simulation is
multiplied by a factor of 0.55, which is approximately the square of
the bias factor, $b=0.75$. Hence, either the galaxies are anti-biased
with respect to the matter clustering, or, the normalization of used
for the n-body simulations should have been approximately
$\sigma_8\simeq 0.75$.  The observed $K$-band selected evolution
appears to be too slow with redshift to readily agree with the
$\Omega=1$ predictions, although the formal level of exclusion is
strongly dependent on the low redshift normalization.  The second
conclusion to be drawn is that the amplitude of the correlations
increases as the color used to select the objects becomes redder and
there is weak evidence that $\gamma$ is also steeper in the $K$
sample.  

\section{Conclusions}

The fitted correlation length of luminous $K$-selected galaxies over the
redshift range 0.2 to 1.2 is substantially stronger than that found for
optically selected samples, about a factor of two in the amplitude,
$r_0^\gamma$.  The galaxy correlation amplitude is measured at a mean
$z\simeq1.39$ as $r_0=1.0e^{\pm0.2}\hmpc$ (formally an upper limit,
but deficient in the more strongly clustered faint red galaxies),
$r_0=1.4e^{\pm0.15}\hmpc$ at $z\simeq0.97$, $2.0e^{\pm0.15}\hmpc$ at
$z\simeq0.62$, and $2.9e^{\pm0.12}\hmpc$ at $z\simeq 0.34$. Together
these give a clustering $\epsilon\simeq0.2\pm0.5$.

The red galaxies are about a factor of 5 more correlated than the blue
galaxies, which have $r_0\simeq0.9e^{\pm0.22}\hmpc$.  These blue
galaxies have a mean equivalent width in the \oii\ line of
41\AA. Together this can be taken as strong evidence that the faint
blue galaxies are an intrinsically weakly correlated population
(therefore likely low mass) with a high star formation rate that
brightens them into the range of much more strongly correlated red
galaxies. These galaxies are so weakly correlated that for our
measured growth of correlations, $\epsilon\simeq0.2\pm0.5$, they would
grow to a current epoch correlation length of $r_0\simeq 2\hmpc$. This
correlation length is less than that measured in any galaxy population
at low redshift (\cite{apm}), so these faint blue galaxies cannot by
themselves make a significant contribution to the current epoch galaxy
population.

Overall the $K$ selected galaxy correlation evolution is somewhat too
slow with redshift to be easily consistent with the evolution of the
matter correlation function for $\Omega_0\simeq1$.  A $\sigma_8\simeq
0.8$ and $\Omega\simeq0.2-0.3$ would describe both the amplitude and
its evolution, if these galaxies are tracing the matter clustering.
To further test models of correlation evolution requires large
datasets of precision comparable to that available in current low
redshift surveys with good control over population changes with
redshift.

\acknowledgments
We thank the referee for constructive criticism of an earlier version
of this paper. RGC acknowledges the financial support of an NSERC
grant.  This work was partly based on observations obtained with the
NASA/ESA Hubble Space Telescope.  Research on the Hawaii Survey fields
was supported by the State of Hawaii and by NASA through grants
GO-5399.01-93A, GO-5922.01-94A, and GO-6626.01-95A from the Space
Telescope Science Institute, which is operated by AURA, Inc., under
NASA contract NAS5-26555.

\clearpage

\figcaption{
The combined redshift distribution of the 248 galaxies with measured
redshifts in the $B,I,K$ magnitude-selected sample of the SSA13 and SSA22
fields. The bins have $\Delta z=0.01$. 
\label{fig:nz}}
\begin{figure}[h] \epsscale{1.0}\figurenum{1}\plotone{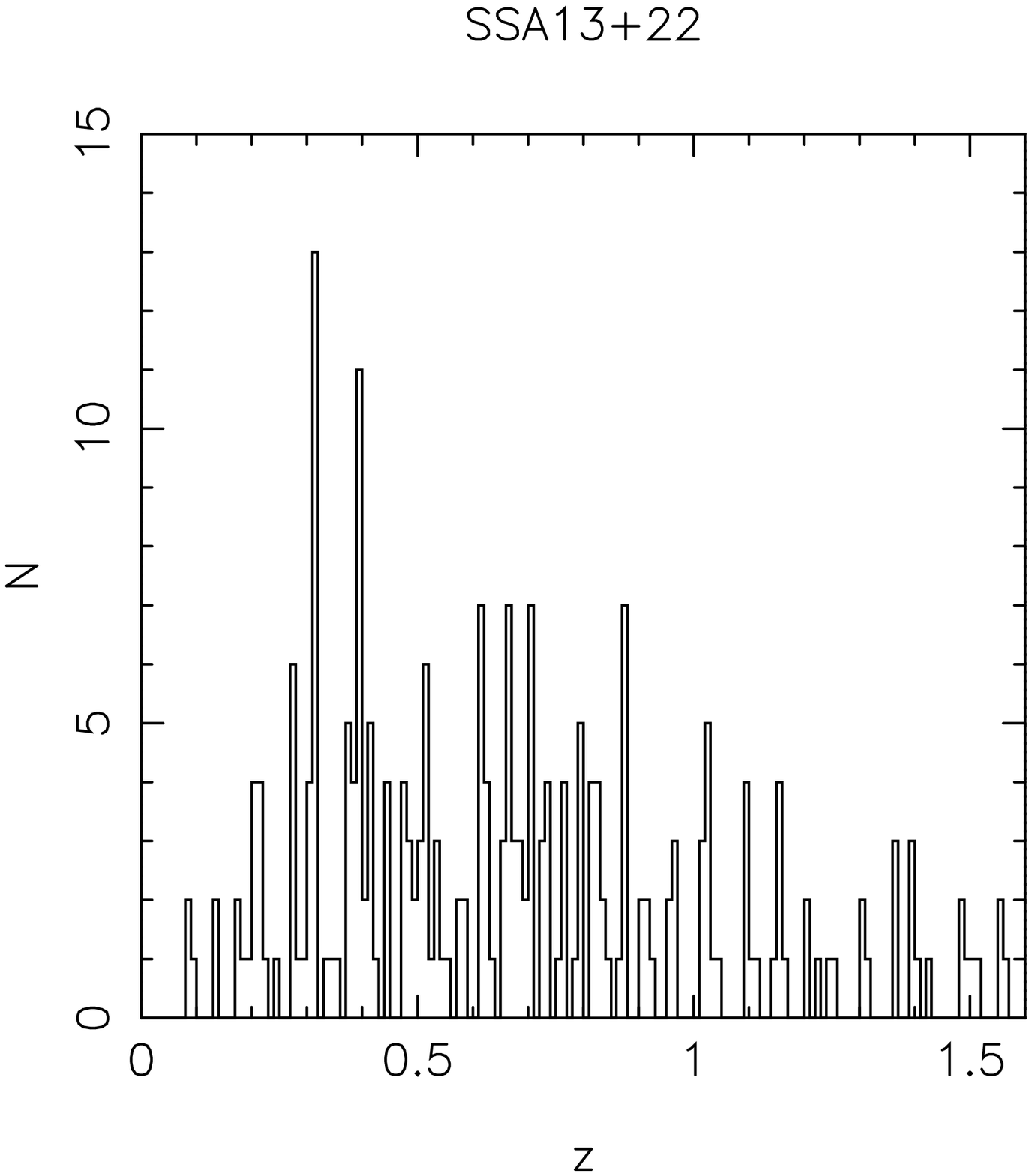} 
\caption{
}\end{figure}
  
\figcaption{
The positions of $K$-detected objects on the sky in SSA13 and SSA22.
The small ticks are at intervals of 20\arcsec\ in RA and
10\arcsec\ in Dec.  Symbol area is proportional to $m_K$. Galaxies
are shown as squares, stars as crosses, objects without confident 
redshift identifications as triangles, and unobserved objects as plus
signs. A larger fraction of objects in the SSA22 field are stars due
to its lower galactic latitude.
\label{fig:xy}}
\begin{figure}[h] \epsscale{0.4}\figurenum{2}\plotone{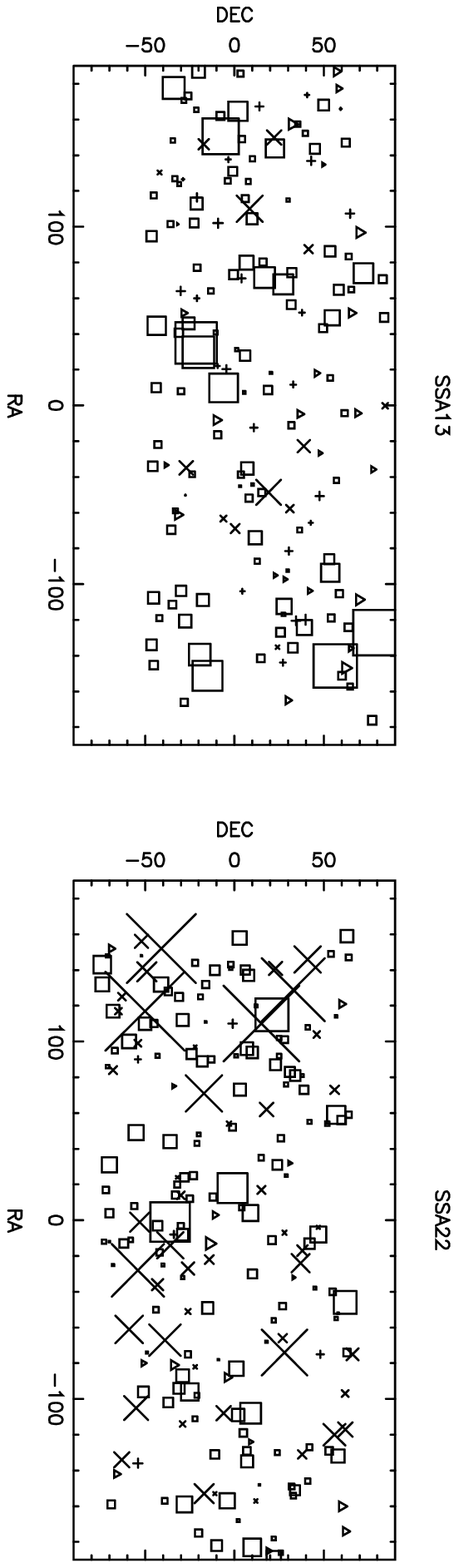} 
\caption{
}\end{figure}

\figcaption{
The redshift vs projected proper distance from field center in the RA 
direction, calculated for $q_0=0.1$ for the entire sample. The area
of each circle is proportional to the object's $K$-band luminosity. The
size for a galaxy with $M_K=-23.5$ is shown near the bottom of the plots 
at $z=1.5$.
\label{fig:pie}}
\begin{figure}[h] \epsscale{0.6}\figurenum{3}\plotone{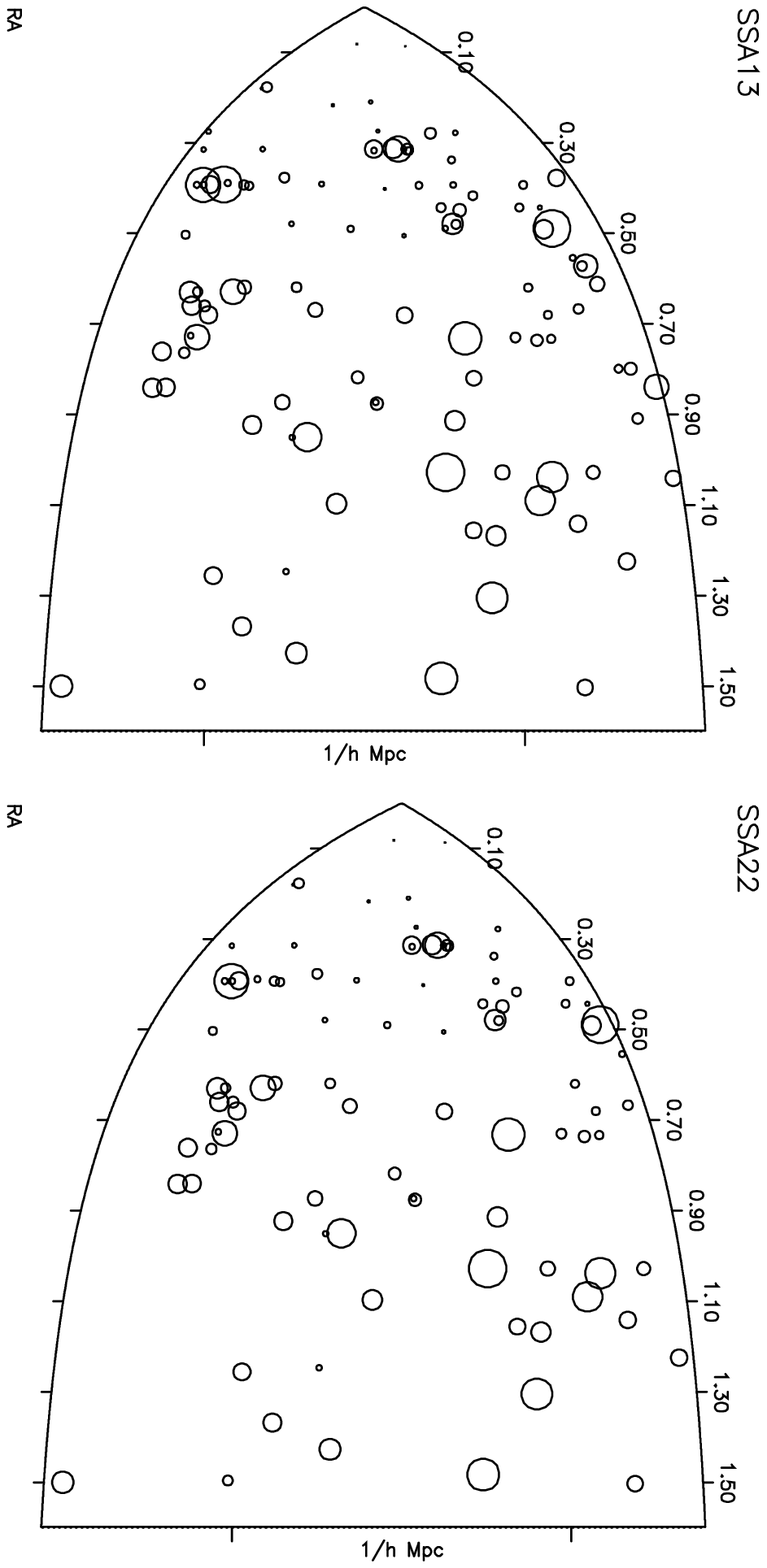} 
\caption{
}\end{figure}

\figcaption{
The angular correlation of the photometric sample (including stars).
The two points at largest separation have negative values, which are not
statistically significant. The indicated $1\sigma$ errors are from a 
bootstrap analysis with 100 resamplings.
\label{fig:aw}}
\begin{figure}[h] \epsscale{1.0}\figurenum{4}\plotone{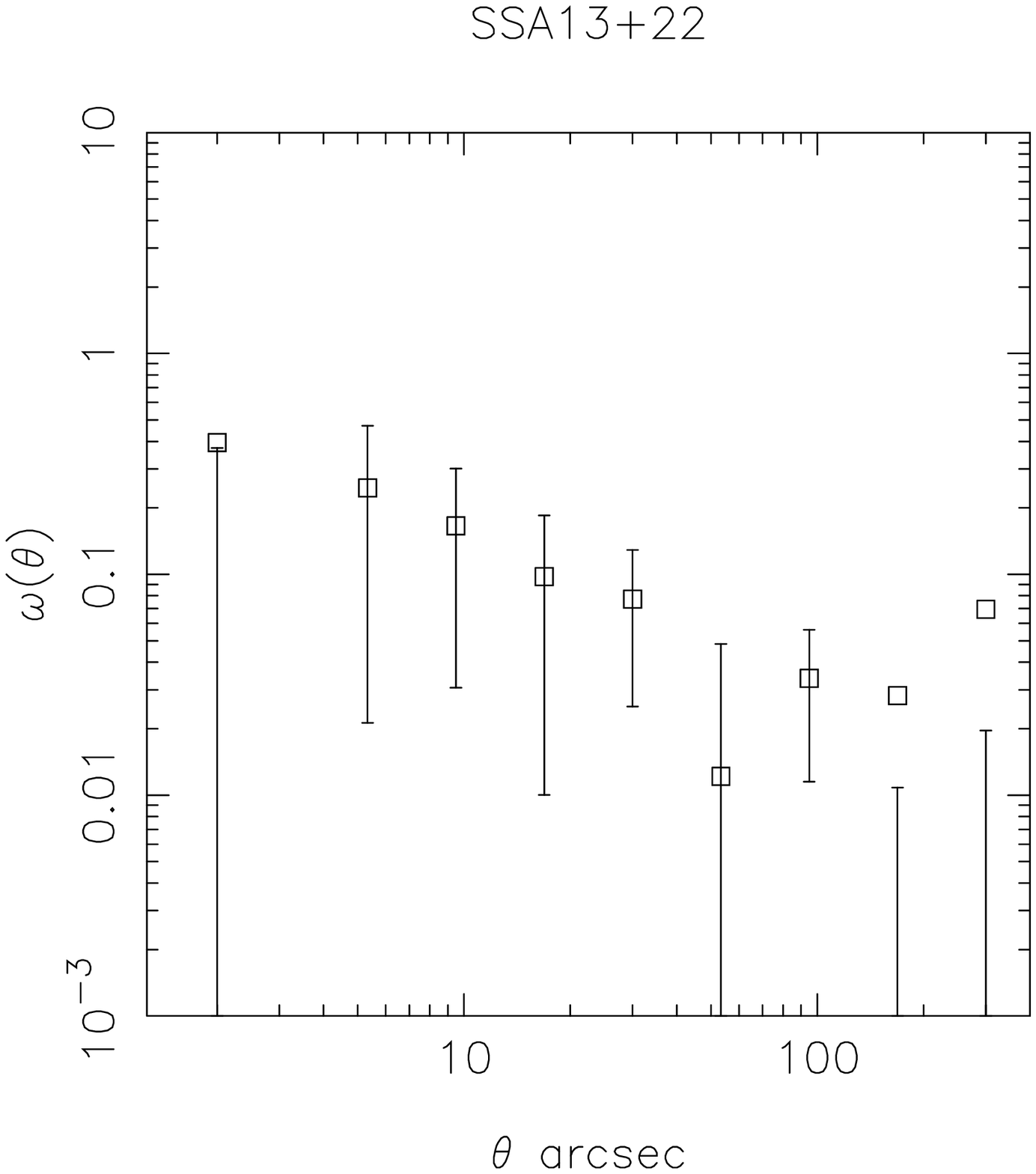} 
\caption{
}\end{figure}  

\figcaption{
The projected real space correlation function for redshift subsamples at 
$0.2\le z \le0.4$ (top-left panel), $0.4\le z \le 0.8$ (top-right panel),
$0.8\le z\le 1.2$ (right panel), and $1.2
\le z \le 1.6$.  The $1\sigma$ bootstap errors (narrow error
flags) indicate that the high redshift correlation is an upper limit,
although the Poisson error bars (wide error flags) indicate a
significant result.  The high redshift subsample is deficient in faint
red galaxies, which are expected to be strongly correlated.  The point
at 0.01\hmpc\ corresponds to an angle of about 2\arcsec\ where galaxy
images overlap and is not used to calculate the correlation
length. All errors are from a bootstrap analysis and are considerably
larger than Poisson estimates.
\label{fig:wp}}
\begin{figure}[h] \epsscale{1.0}\figurenum{5}\plotone{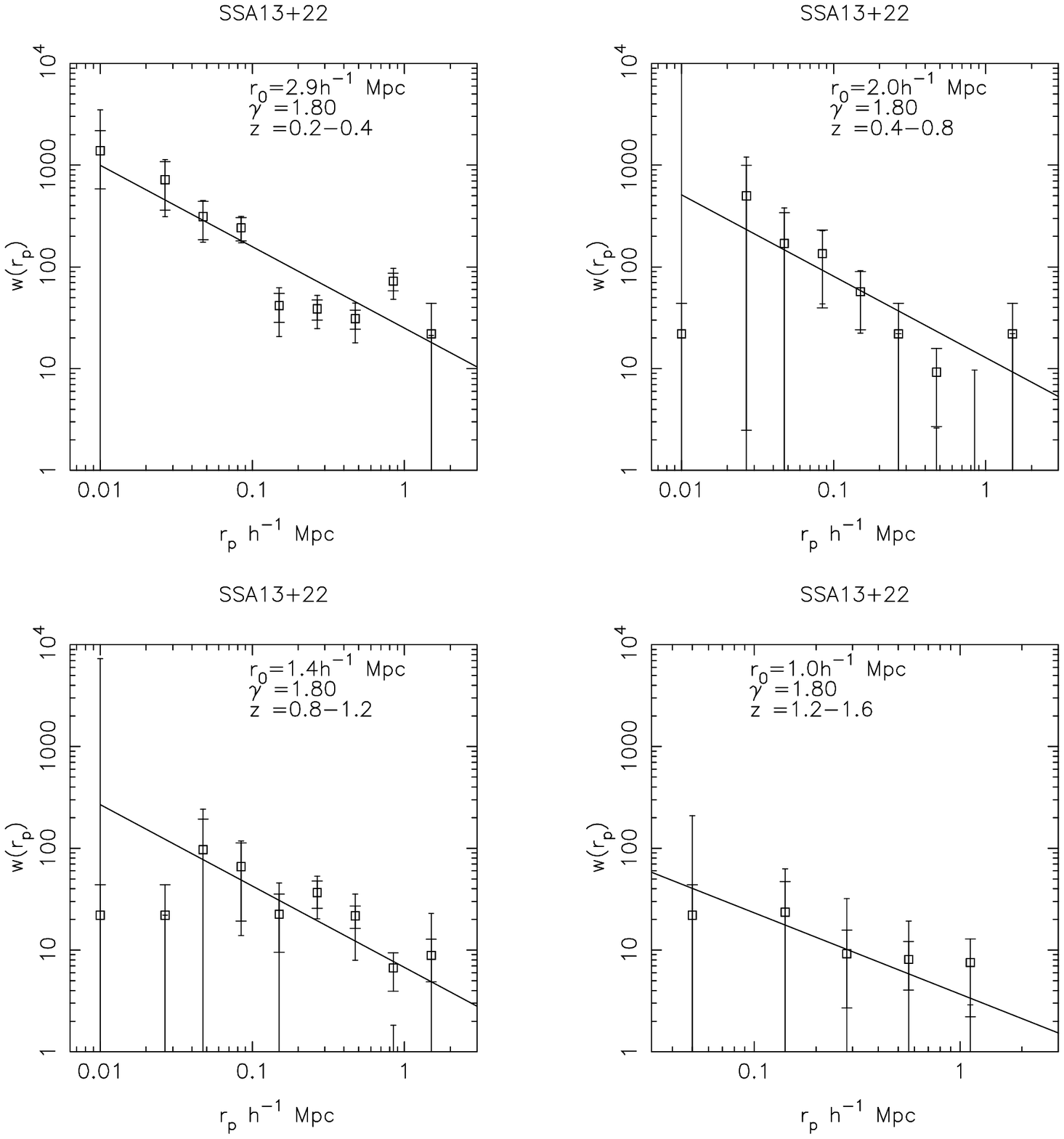} 
\caption{
}\end{figure}

\figcaption{
The projected real space correlation function for the blue and red
subsamples over $0.3\le z \le0.9$. Bootstrap errors are shown.
\label{fig:wc}}
\begin{figure}[h] \epsscale{0.6}\figurenum{6}\plotone{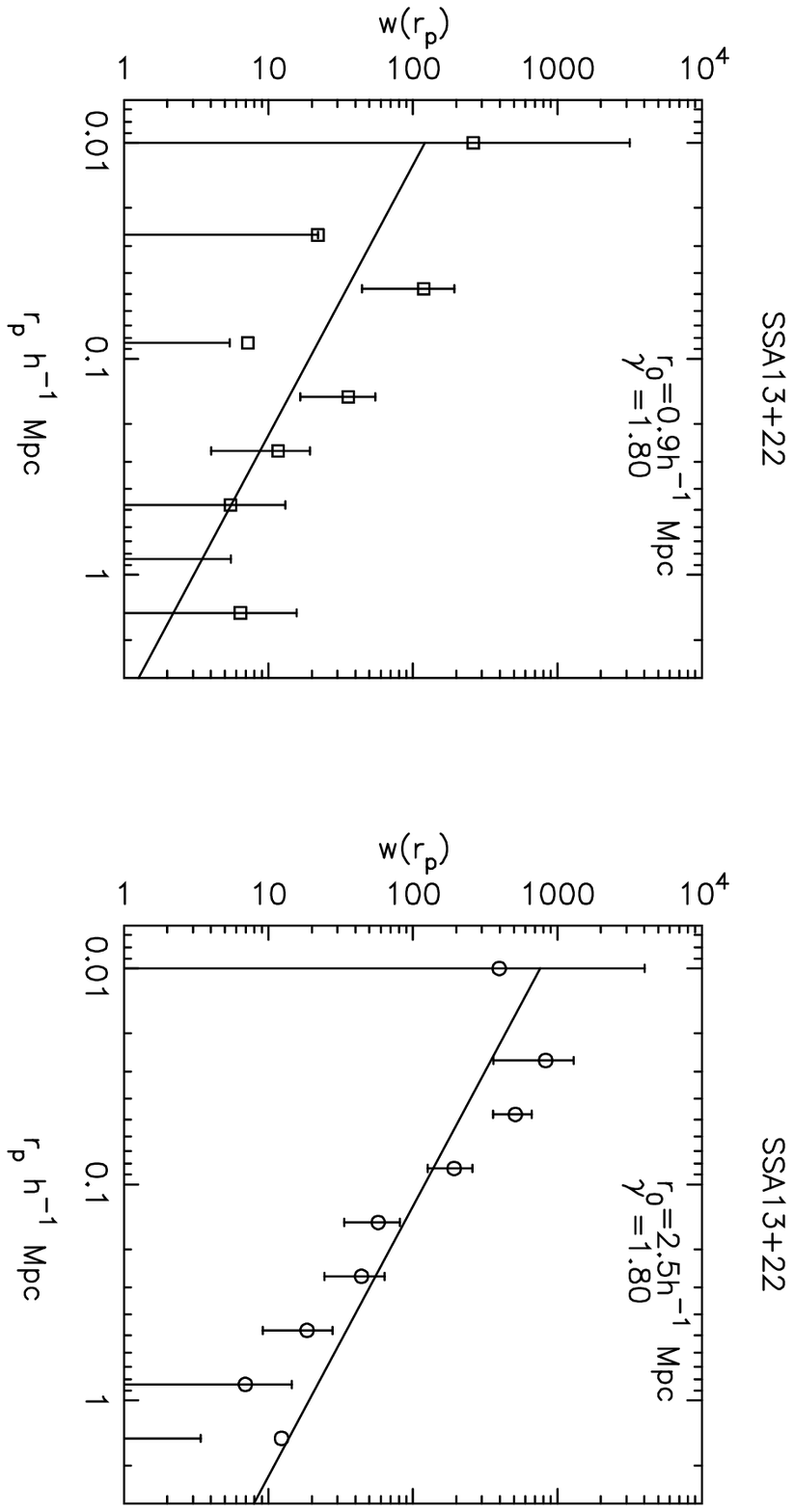} 
\caption{
}\end{figure}
  
\figcaption{
The projected real space cross-correlation function of the $M_K\ge
-23.5$ subsample with the $M_K\le -23.5$ subsample.  The line shows
the auto-correlation of high luminosity galaxies over the same
redshift range.  Bootstrap errors are shown.
\label{fig:w2}}
\begin{figure}[h] \epsscale{1.0}\figurenum{7}\plotone{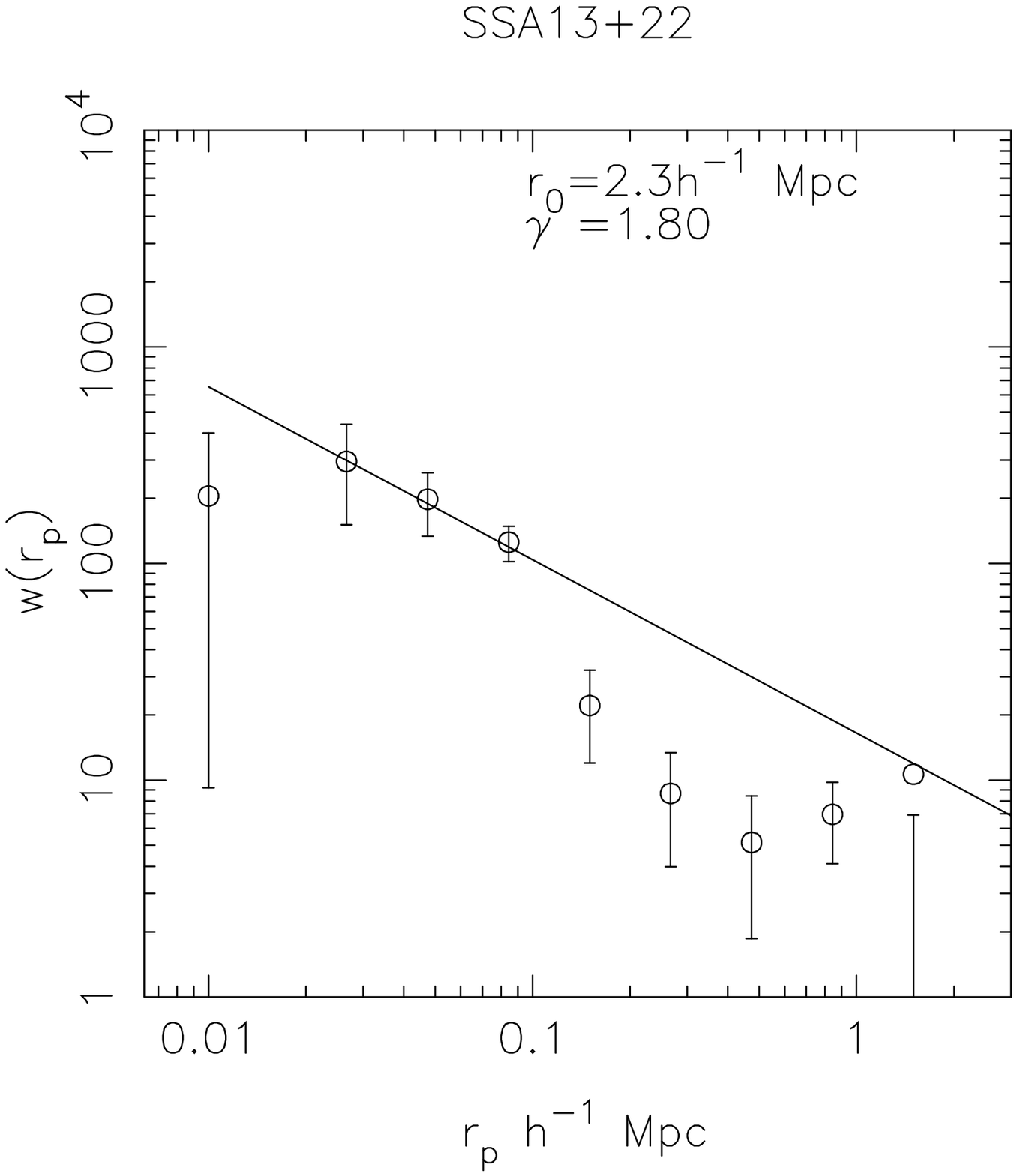} 
\caption{
}\end{figure}

\figcaption{
The evolution of the nonlinear amplitude of the correlation function,
$r_0^\gamma$.  The data are compared with results from other surveys
in the literature (see Section 6 for details), with all measurements
adjusted to $q_0=0.1$.  The results from this paper are shown as
filled circles, with superposed crosses for the blue and red
subsamples.  The solid, dashed, and dotted lines are based on power
law fits to the $\xi(r|z)$ measured in n-body simulations of CDM
universes for different assumed values of $\Omega$ and $\Lambda$,
renormalized to pass through the LCRS measurement at low
redshift. Note that correlations of relatively red selected galaxies
are always larger than those of blue selected galaxies. Bootstrap
errors are shown. The Keck and BGFS samples are $K$ selected.  CNOC
and LCRS are $r$ selected, and CFRS is $I$ selected. The APM is blue
selected and the IRAS sample is dominated by relatively blue galaxies
which contain warm dust.
\label{fig:ev}}
\begin{figure}[h] \epsscale{1.0}\figurenum{8}\plotone{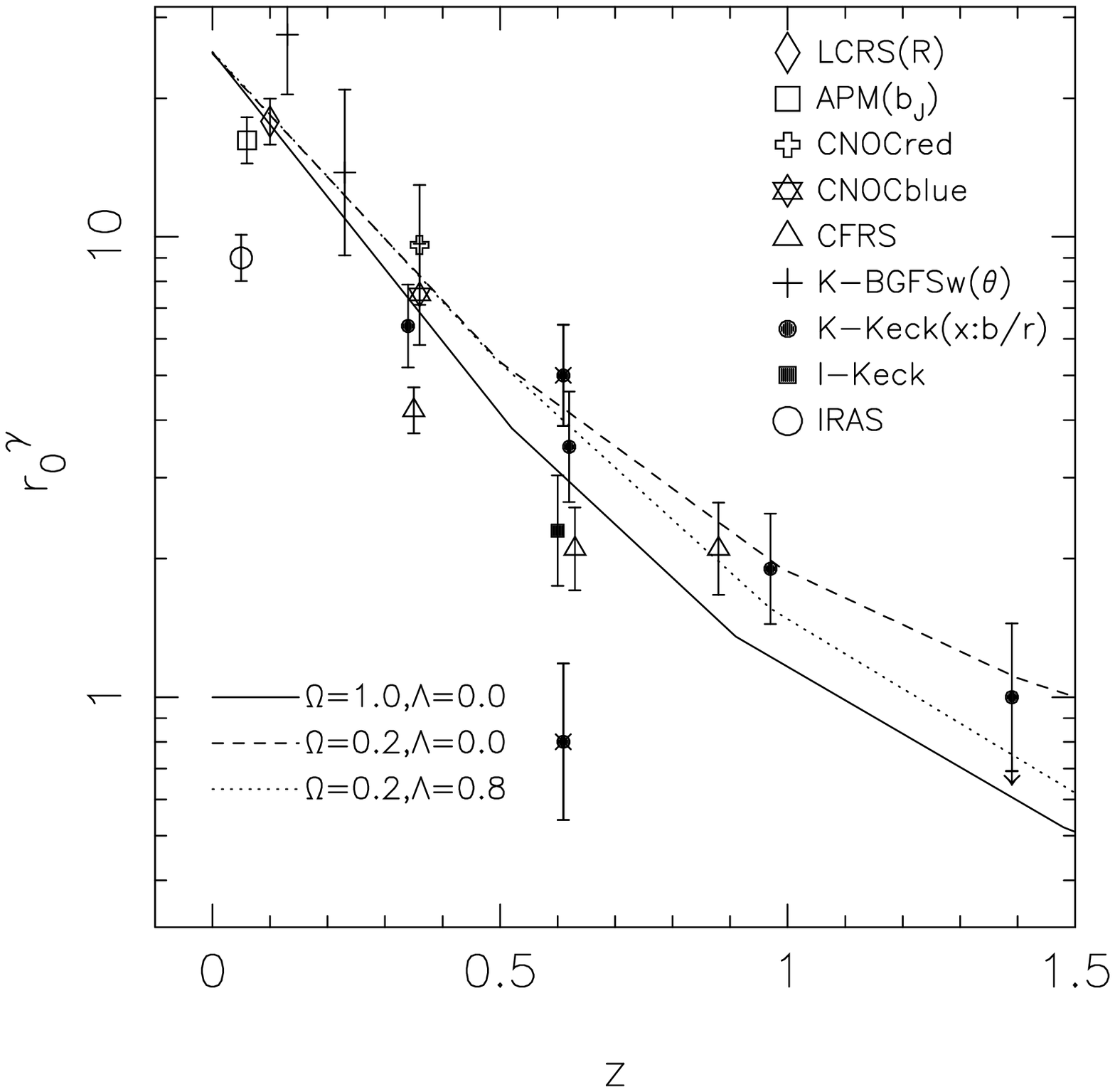} 
\caption{
}\end{figure}

\end{document}